\documentclass[conference]{IEEEtran}
\IEEEoverridecommandlockouts
\usepackage{verbatim}
\usepackage{graphicx, color}
\usepackage{hyperref}
\hypersetup{
    colorlinks=true,
    citecolor=,
    linkcolor=,
    urlcolor=cyan,
}

\title{Artifact: Measuring and Mitigating Gaps in Structural Testing}

\author{\IEEEauthorblockN{Soneya Binta Hossain}
\IEEEauthorblockA{\textit{Department of Computer Science} \\
\textit{University of Virginia}\\
Charlottesville, USA \\
sh7hv@virginia.edu}
\and
\IEEEauthorblockN{Matthew B. Dwyer}
\IEEEauthorblockA{\textit{Department of Computer Science} \\
\textit{University of Virginia}\\
Charlottesville, USA \\
matthewbdwyer@virginia.edu}
\and

\IEEEauthorblockN{Sebastian Elbaum}
\IEEEauthorblockA{\textit{Department of Computer Science} \\
\textit{University of Virginia}\\
selbaum@virginia.edu}

\and
\IEEEauthorblockN{\centerline{Anh Nguyen-Tuong}}
\IEEEauthorblockA{\centerline{\textit{Department of Computer Science}} \\
\textit{University of Virginia}\\
an7s@virginia.edu}
}

\begin{document}

\maketitle
\begin{abstract}
The artifact used for evaluating the experimental results of \textit{Measuring and Mitigating Gaps in Structural Testing} is publicly \textit{available} on GitHub, Software Heritage and figshare, and is \textit{reusable}. The artifact consists of necessary data, tools, scripts, and detailed documentation for running the experiments and reproducing the results shown in the paper. We have also provided a \textit{VirtualBox VM} image allowing users to quickly setup and reproduce the results. Users are expected to be familiar using the VirtualBox software and Linux platform for evaluating or reusing the artifact.
\end{abstract}


\section{Introduction}
Our research aims to measure and mitigate \textit{gaps} in structural testing, i.e., the portion of code structures requiring more testing. For measuring coverage gaps, we have proposed host checked coverage (HCC), an extension of checked coverage \cite{schuler2011assessing}. The gap is calculated as the percentage points (\textit{pp}) 
difference between regular code coverage and host checked coverage. Our study shows that the gap is \textit{strongly} and \textit{negatively} correlated with the fault-detection effectiveness of a test suite. To mitigate gaps, we have proposed a \textit{recommender} method that suggests ways to reduce gaps by enriching the test suite with additional test oracles. Reducing gaps yields improvement in fault-detection effectiveness. 

This abstract specifies the details of the artifact, which is available on GitHub at \href{https://github.com/soneyahossain/hcc-gap-recommender}{https://github.com/soneyahossain/hcc-gap-recommender} (also available on Software Heritage \href{https://archive.softwareheritage.org/swh:1:dir:3a0a51add2930bb96b8b6ea2cd76a4fe47cda032}{repository}) and on long-term data archival repository figshare at \href{https://doi.org/10.6084/m9.figshare.21950552}{https://doi.org/10.6084/m9.figshare.21950552}. All repositories are public and include all required data, tool source codes, scripts, and detailed instructions to build the tools and run the experiments to reproduce the results presented in the paper. The VirtualBox VM image, README, REQUIREMENTS, STATUS, LICENSE, INSTALL, a copy of the accepted paper can be found in the \href{https://doi.org/10.6084/m9.figshare.21950552}{figshare} repository. 

\begin{comment}
Bash scripts are provided to reproduce RQ-wise experimental results and run a time-limited version of the end-to-end workflow. All necessary scripts are provided for running the whole experiment as well. However, to run the entire study takes several days. The most expensive parts of our study are the computation of program slices and mutation testing.

We also provide a VirtualBox VM image in the \href{https://doi.org/10.6084/m9.figshare.21950552}{figshare} repository with detailed instructions to simplify artifact assessment. The VM comes with a pre-installed, pre-tested execution environment.
\end{comment}

\section{Motivation} 
Regular code coverage \textit{only} measures the percentage of program codes executed by a test suite; therefore, it does not provide any insight into the quality of test oracles. In this research, we have modified and extended the original definition of checked coverage \cite{schuler2011assessing} to identify the under-tested program structures that require more testing. We refer to these under-tested codes as the \textit{coverage gaps}. These gaps indicate the program structures (e.g., statement, object branch) that are executed but not checked by any test oracles. A large-scale study revealed a strong negative correlation between the gap and fault-detection effectiveness of a test suite. 
To close the gap and improve fault detection, 
%The coverage gap is valuable information that can be utilized to add more test assertions to close that gap. 
we have implemented a lightweight static recommender that recommends additional assertions.
%to improve test suites' fault-detection capability. 

\begin{comment}
\section{Badges Claimed}  
We claim that the artifact is both \textit{available} and \textit{reusable}.
The implementation is publicly available at \href{https://github.com/swa112003/DistributionAwareDNNTesting}{https://github.com/swa112003/DistributionAwareDNNTesting} and its usage is well documented in the \texttt{README} file of the repository. 
Apart from reproducing the experiments used in the paper, the implementation can be easily extended for testing new DNN models. We will follow the steps listed out in artifact evaluation guidelines and provide the required documentation in our submission in preparing the artifact.
\end{comment}

\section{Implementation} 

The artifact implements a time-limited version of the end-to-end workflow of the HCC framework and the research questions answered in our paper \cite{10172745}. The time-limited scripts use fragment of data for quick evaluation 
%and functionality check 
as the entire study takes several days.

For computing HCC, we first record execution traces of a test suite using JavaSlicer \cite{hammacher2008design}, automatically generate slicing criteria and then compute dynamic slices using JavaSlicer. Once all slices are constructed, we compute statement checked coverage (SCC) and object-branch checked coverage (OBCC) using our implemented tools. Finally, we calculate the coverage gap from the HCC and regular code coverage. Pre-built jar files are located in the \texttt{lib} directory, and source codes for all tools are located in the \texttt{hcctools} directory. \texttt{../experiments/scripts/smoke-tests.sh} executes the above steps to demonstrate that our artifact is \textit{functional} and \textit{reusable}.  

All necessary scripts to generate intermediate results, such as building a particular Java subject, generating regular coverage, recording trace, computing slice, SCC, OBCC, gap, recommendation, manipulating test suite to vary gap, and running mutation tests, can be found in the \texttt{experiments/scripts} directory.

For evaluating specific research questions, we have provided bash scripts. For example, \textit{experiments/scripts/rq1.sh} computes statement and object branch coverage, statement checked coverage (SCC) and object branch checked coverage (OBCC). Results are stored in a .csv file which can be used to compute the coverage gaps shown in RQ1 (Table II). 

For evaluating RQ2, we have provided \textit{experiments/scripts/rq2.sh}, which runs 
\textit{hcc-gap-recommender/hcctools/testsuitegen/} tool to generate several versions of test suites with varied coverage gaps. Then, mutation testing \cite{coles2016pit} is performed on each test suite to estimate their fault-detection effectiveness. 

For RQ3, \textit{rq3.sh} is used to reproduce the results in Table III. The outputs are stored in a .csv file with the top-k scores. 

For RQ4, we have computed SCC for two different versions of joda-time chronology test suites, one is the default and the other is enriched with  oracles suggested by the recommender. 

More details about these commands and the repository structure can be found in  \href{https://github.com/soneyahossain/hcc-gap-recommender/blob/main/README.md}{README}. 

\section{Running the artifact}  
\begin{comment}
Reviewers should be familiar with running Python based machine learning scripts and using Linux operating system. Since it is time consuming to run all the experiments in the paper, we provide a VirtualBox VM image with fully configured execution environment and a subset of the experiments from the paper. The reviewers are expected to have knowledge on using VirtualBox software for running these tests. 
\end{comment}

Users should be familiar with running Java-based commands, bash scripts, VirtualBox software, and using the Linux operating system. As program slicing takes a comparatively long time, we have provided a bash script, \textit{experiments/scripts/smoke-tests.sh} to run the HCC framework end-to-end, including tracing, slicing, HCC, gap, and recommendation computation for a small fraction of data. 

The VirtualBox VM has a fully configured execution environment and all relevant tools, libraries, and environment variables are already set up. However, when running on personal machines, one must follow the instructions in the \href{https://github.com/soneyahossain/hcc-gap-recommender/blob/main/README.md}{README} to setup the environment.

Once setup correctly, run the end-to-end smoke-tests using the following commands:

\begin{verbatim}
cd $HCC_EXPERIMENTS/scripts
./smoke-tests.sh
\end{verbatim}
This will produce the following output:
\begin{verbatim}
************************************
Smoke tests for end-to-end workflow
************************************
Verify ability to generate:
   - statement coverage via clover
   - object branch coverage via JaCoCo
   - traces via JavaSlicer
   - slices via JavaSlicer
   - statement checked coverage (SCC)
   - object branch coverage (OBCC)
   - recommendations via recommender
.
Trace file generated: OK
........
Slice file(s) generated: OK
...........
SCC computed: OK
.........
OBCC computed: OK
............
Recommender ran successfully: OK 
\end{verbatim}

The \texttt{README} file in the GitHub repository provides the detailed output log. Upon successful completion, all results will be stored in the \textit{experiments/hcc\_results/commons-cli-limited/} directory. All .csv files starting with ``scc'' prefix contain statement checked coverage results and scc.csv contains the overall summary. Similarly, all .csv files starting with ``obcc'' prefix contain object branch checked coverage results and obcc.csv consists of the overall summary. The recommender results will be stored in \textit{experiments/hcc\_results/commons-cli-limited/evaluator/} directory and summary.csv contains the overall summary. 

For running the experiment corresponding to a research question, users should go to the \textit{experiments/scripts/} directory using the \textit{rq1.sh}, \textit{rq2.sh}, \textit{rq3.sh}, or \textit{rq4.sh} bash scripts.
%and run ``./rqN.sh'', where N is 1,2,3, or 4. 
Details on what to expect after successful completion is provided in \href{https://github.com/soneyahossain/hcc-gap-recommender/blob/main/README.md}{README}. 

We will keep the GitHub repository up to date as the HCC framework evolves in the future. 
The current VirtualBox VM image captures the artifact at the time of publication and provides a fully configured execution environment.

\section{Requirements}   
All tools associated with this artifact require a platform with Linux operating system, JDK 1.7 and 1.8, and maven 3.6.3. JavaSlicer can be installed from \url{https://github.com/backes/javaslicer}, and we have also provided pre-built jar files in our artifact repository. We tested on an Ubuntu 20.04.3 LTS platform. For user convenience, we suggest using the VirtualBox VM image with a fully configured execution environment and following the instructions provided in the \href{https://github.com/soneyahossain/hcc-gap-recommender/blob/main/README.md}{README} to run experiments. 

\section{Acknowledgements}
This material is based in part upon work supported by the DARPA ARCOS program under contract FA8750-20-C-0507, by the Air Force Office of Scientific Research under award number FA9550-21-0164, and by Lockheed Martin Advanced Technology Laboratories.

\bibliographystyle{abbrv}
\bibliography{a}

\begin{thebibliography}{1}

\bibitem{coles2016pit}
H.~Coles, T.~Laurent, C.~Henard, M.~Papadakis, and A.~Ventresque.
\newblock Pit: a practical mutation testing tool for java.
\newblock In {\em Proceedings of the 25th international symposium on software
  testing and analysis}, pages 449--452, 2016.

\bibitem{hammacher2008design}
C.~Hammacher.
\newblock Design and implementation of an efficient dynamic slicer for java.
\newblock {\em Bachelor's Thesis}, 2008.

\bibitem{10172745}
S.~B. Hossain, M.~B. Dwyer, S.~Elbaum, and A.~Nguyen-Tuong.
\newblock Measuring and mitigating gaps in structural testing.
\newblock In {\em 2023 IEEE/ACM 45th International Conference on Software
  Engineering (ICSE)}, pages 1712--1723, 2023.

\bibitem{schuler2011assessing}
D.~Schuler and A.~Zeller.
\newblock Assessing oracle quality with checked coverage.
\newblock In {\em 2011 Fourth IEEE International Conference on Software
  Testing, Verification and Validation}, pages 90--99. IEEE, 2011.

\end{thebibliography}
\end{document}